\documentclass[a4paper]{jpconf}
\usepackage{graphicx}
\usepackage[hidelinks]{hyperref}
\usepackage{color}
\usepackage{amsmath,amsfonts,amssymb,bm}

\usepackage{lipsum}

\usepackage{fancyhdr}
\def\beq{\begin{equation}}
\def\eeq{\end{equation}}
\def\beqn{\begin{eqnarray}}
\def\eeqn{\end{eqnarray}}
\def\beqs{\begin{subequations}}
\def\eeqs{\end{subequations}}

\def\P {{\bf P}}

\def\bi {\mbox{\boldmath $i$}}

\def\P {{\bf P}}

\begin{document}

\title{Correlation Effects in a Trapped Bose-Fermi Mixture: Exact Results}

\author{O~E~Alon$^{1,2}$ and L~S~Cederbaum$^{3}$}
\address{$^{1}$ Department of Physics, University of Haifa, 3498838 Haifa, Israel}
\address{$^{2}$ Haifa Research Center for Theoretical Physics and Astrophysics, University of Haifa, 3498838 Haifa, Israel}
\address{$^{3}$ Theoretical Chemistry, Physical Chemistry Institute, Heidelberg University, D-69120 Heidelberg, Germany}

\ead{ofir@research.haifa.ac.il}

\begin{abstract}
Many-body properties of a fermionic impurity embedded in a Bose-Einstein condensate
are analyzed analytically using a solvable model,
the harmonic-interaction model for Bose-Fermi mixtures.
The one-particle and two-particle densities, reduced density matrices, and correlation functions of the fermions and bosons,
both in position and momentum spaces, are prescribed in closed form.
The various coherence lengths are analyzed.
We show that the first-order coherence lengths
in position and momentum spaces are equal
whereas the second-order quantities can differ substantially.
Illustrative examples where the sole interaction is between the impurity and the condensate are presented.
Implications are briefly discussed.    
\end{abstract}

\section{Introduction}\label{SEC_INTRO}

Ultracold atomic Bose-Fermi mixtures have generated a vast and extensive experimental and theoretical research
for more than two decades,
see, e.g., Refs.~[1-32].
There are endless combinations, if one adds together intraspecies and interspecies interactions, 
numbers of particles of each species and their masses, and trapping potentials,
to the inherent different nature of the Bose and Fermi statistics.
On the theory side,
this wealth offered by the ultracold realm makes solvable models rather scarce,
especially in traps.
The utility of exactly-solvable models of quantum many-particle systems is indispensable
as tools to emulate such systems and to gain instincts on realistic scenarios.
These are additional to the mere mathematical aesthetics which usually accompanies such models.  

Harmonic-interaction models have amply been used to emulate many-particle systems for many years
[33-52].
Another usage of harmonic-interaction models is to benchmark
many-body theories and their numerical implementations
[53-57].
The harmonic-interaction models have been mainly used
for single-species bosons, single-species fermions, and Bose-Bose mixtures so far.
In the present work we wish to add another brick to the building of harmonic-interaction models,
and treat Bose-Fermi mixtures with it.
Explicitly,
we investigate analytically first-order and second-order correlations
of a trapped fermionic impurity in a Bose-Einstein condensate both in position and momentum spaces.
We build on derivations and results using the harmonic-interaction model for Bose-Fermi mixtures
which will be reported in an extended format elsewhere \cite{arXiv_2023}.

\section{Theory}\label{SEC_THEORY}

The Bose-Fermi Hamiltonian within the harmonic-interaction model reads
\beqn\label{BF_HAM_LP}
& & \hat H(x_1,\ldots,x_{N_F},y_1,\ldots,y_{N_B}) = \nonumber \\
& & = \sum_{j=1}^{N_F} \left( -\frac{1}{2m_F} \frac{\partial^2}{\partial x_j^2} + \frac{1}{2} m_F\omega^2 x_j^2 \right)  
+ \sum_{j=1}^{N_B} \left( -\frac{1}{2m_B} \frac{\partial^2}{\partial y_j^2} +
\frac{1}{2} m_B\omega^2 y_j^2 \right) + \nonumber \\
& & + 
\lambda_F \sum_{1 \le j < k}^{N_F} \left(x_j-x_k\right)^2 + 
\lambda_B \sum_{1 \le j < k}^{N_B} \left(y_j-y_k\right)^2 +  
\lambda_{BF} \sum_{j=1}^{N_F} \sum_{k=1}^{N_B} \left(x_j-y_k\right)^2, \
\eeqn
where $N_F, N_B$ are the numbers of fermions and bosons,
$m_F, m_B$ their masses,
and $\lambda_F, \lambda_B$ the intraspecies
and $\lambda_{BF}$ the interspecies interactions strengths, respectively.
Herein, the fermions are polarized.

Diagonalizing (\ref{BF_HAM_LP}) leads to four distinct eigenfrequencies, also see the Bose-Bose mixture \cite{EPJD_2014,JPA_2017}
\beqn\label{BF_FREQ_LP}
& & \Omega_{F} = \sqrt{\omega^2 + \frac{2}{m_F}\left(N_F\lambda_F+N_B\lambda_{BF}\right)}, \qquad
\Omega_{B} = \sqrt{\omega^2 + \frac{2}{m_B}\left(N_B\lambda_B+N_F\lambda_{BF}\right)}, \nonumber \\
& & \Omega_{BF} = \sqrt{\omega^2 + 2\left(\frac{N_B}{m_F}+\frac{N_F}{m_B}\right)\lambda_{BF}}, \qquad \omega. \
\eeqn
The ground-state energy of $N_F$ fermions and $N_B$ bosons is \cite{arXiv_2023}
\beqn\label{BF_Energy_LP}
& &
E = \frac{1}{2} \Bigg[\left(N_F^2-1\right)\sqrt{\omega^2 + \frac{2}{m_F}(N_F\lambda_F+N_B\lambda_{BF})} + \nonumber \\
& & + \left(N_B-1\right) \sqrt{\omega^2 + \frac{2}{m_B}(N_B\lambda_B+N_F\lambda_{BF})} +
\sqrt{\omega^2 + 2\left(\frac{N_B}{m_F}+\frac{N_F}{m_B}\right)\lambda_{BF}} + \omega\Bigg],
\eeqn
and the wavefunction in Cartesian coordinates reads
\beqn\label{BF_PSI_CAR_LP}
& & \Psi(x_1,\ldots,x_{N_F},y_1,\ldots,y_{N_B}) = \nonumber \\
& & = {\mathcal{N}}
\left(\frac{m_F\Omega_F}{\pi}\right)^{\frac{N_F^2-1}{4}}
\left(\frac{m_B\Omega_B}{\pi}\right)^{\frac{N_B-1}{4}}
\left(\frac{M_{BF}\Omega_{BF}}{\pi}\right)^{\frac{1}{4}}
\left(\frac{M\omega}{\pi}\right)^{\frac{1}{4}}
\times \nonumber \\
& & \times
\left[\prod_{1 \le j < k \le N_F} \left(x_j-x_k\right)\right]
e^{-\frac{\alpha_F}{2} \sum_{j=1}^{N_F} x_j^2 - \beta_F \sum_{1 \le j < k}^{N_F} x_j x_k}
e^{-\frac{\alpha_B}{2} \sum_{j=1}^{N_B} y_j^2 - \beta_B \sum_{1 \le j < k}^{N_B} y_j y_k}
\times \nonumber \\
& & \times e^{+\gamma \sum_{j=1}^{N_F} \sum_{k=1}^{N_B} x_j y_k}, \
\eeqn
where
$\prod_{1 \le j < k \le N_F} \left(x_j-x_k\right) = V(x_1,\ldots,x_{N_F})$
is the Vandermonde determinant
and ${\mathcal{N}}$ is a dimensionless normalization
constant that appears in the derivation due to the relative
Jacoby coordinates of the fermions \cite{arXiv_2023}.
The center-of-mass and relative center-of-mass degrees-of-freedom
have the masses
$M=m_FN_F+m_BN_B$ and $M_{BF}=\frac{m_Fn_B}{M}$, respectively.

There are five coefficients in the wavefunction, also see \cite{JPA_2017},
\beqn\label{BF_PSI_CAR_PAR_LP}
& & \alpha_F = m_F\Omega_F \left[1 + \frac{1}{N_F}\left(\frac{m_FN_F}{M}\frac{\omega}{\Omega_F}+\frac{m_BN_B}{M}\frac{\Omega_{BF}}{\Omega_F}-1\right)\right], \qquad \beta_F = \alpha_F - m_F\Omega_F,
\nonumber \\
& & \alpha_B = m_B\Omega_B \left[1 + \frac{1}{N_B}\left(\frac{m_BN_B}{M}\frac{\omega}{\Omega_B}+\frac{m_FN_F}{M}\frac{\Omega_{BF}}{\Omega_B}-1\right)\right], \qquad \beta_B = \alpha_B - m_B\Omega_B,
\nonumber \\
& & \gamma = \frac{m_Fm_B}{M}(\Omega_{BF}-\omega), \
\eeqn
which are used to express and evaluate all the correlation properties in the Bose-Fermi mixture.
Explicitly, the reduced one-particle and two-particle density matrices \cite{RDM1,RDM2} and 
Glauber's normalized first-order and second-order correlation functions
[61-63]
of the fermions and bosons,
both in position and momentum spaces, are put forward in closed form and investigated.

\subsection{Position and momentum correlation functions of the bosons}\label{SEC_B}

Consider an impurity consisting of $N_F=2$ fermions
embedded in a Bose-Einstein condensate made of $N_B$ bosons.
Two fermions in the impurity already
lead to rich results. 
Of course, this is the minimal number of indistinguishable particles to exhibit
the antisymmetry of the wavefunction (\ref{BF_PSI_CAR_LP}) under their exchange.
The results for fermions obtained below are hence
explicit to two fermions,
although some of them hold for a general
number $N_F$ of fermions.
It is instructive to analyze the condensate first and than the impurity,
since the bosons' correlation functions are relatively simpler than the fermions' ones.

The reduced one-particle density matrix and the density of the bosons in position space
are integrated from the all-particle density matrix and read \cite{JPA_2017,arXiv_2023}
\beqn\label{B_1RDM_DENS_POS_LP}
& & \rho^{(1)}_B(y,y')
= N_B \left(\frac{\alpha_B+C'_{0,1}}{\pi}\right)^{\frac{1}{2}}
e^{- \frac{\alpha_B+C'_{0,1}}{4}\left(y+y'\right)^2}
e^{-\frac{\alpha_B}{4}\left(y-y'\right)^2}, \nonumber \\
& & \rho^{(1)}_B(y)
= N_B \left(\frac{\alpha_B+C'_{0,1}}{\pi}\right)^{\frac{1}{2}}
e^{-\left(\alpha_B+C'_{0,1}\right) y^2}. \
\eeqn
We can now compute explicitly from (\ref{B_1RDM_DENS_POS_LP})
the first-order correlation function in position space,
\beqn\label{B_g1_POS_LP}
& & g_B^{(1)}(y,y') = \frac{\rho^{(1)}_B(y,y')}{\sqrt{\rho^{(1)}_B(y)\rho^{(1)}_B(y')}}
= e^{-\frac{-C'_{0,1}}{4}\left(y-y'\right)^2}, \
\eeqn
where $C'_{0,1}$ is negative and 
given in closed form as a function of the coefficients (\ref{BF_PSI_CAR_PAR_LP}) \cite{JPA_2017,arXiv_2023}, see below.

It is useful to define a dimensionless measure, the first-order coherence length in position space,
as the ratio of the off-diagonal decay length of the first-order correlation function (\ref{B_g1_POS_LP})
and the diagonal decay length of the reduced one-particle density matrix (\ref{B_1RDM_DENS_POS_LP}), see \cite{arXiv_2023},
\beqn\label{l_B_1_POS_LP}
& &
l_B^{(1)} = \sqrt{\frac{\alpha_B+C'_{0,1}}{-C'_{0,1}}} =
\frac{1}{\sqrt{\left(\frac{1+\rho_B^{(1)}}{1-\rho_B^{(1)}}\right)^2-1}} = \\
& & 
= \frac{1}{\sqrt{\left[1 + \frac{1}{N_B}\left(\frac{m_BN_B}{M}\frac{\omega}{\Omega_B}+\frac{m_FN_F}{M}\frac{\Omega_{BF}}{\Omega_B}-1\right)\right]\left[1 + \frac{1}{N_B}\left(\frac{m_BN_B}{M}\frac{\Omega_B}{\omega}+\frac{m_FN_F}{M}\frac{\Omega_B}{\Omega_{BF}}-1\right)\right]-1}}. \nonumber \
\eeqn
The second expression of $l_B^{(1)}$ in (\ref{l_B_1_POS_LP}) is 
because it is possible to express the first-order coherence length using the depleted fraction $0 \le \rho_B^{(1)} < 1$.
The depleted fraction is given in closed-form using the above coefficients,
$\rho_B^{(1)} = \frac{\sqrt{\alpha_B}-\sqrt{\alpha_B+C'_{0,1}}}{\sqrt{\alpha_B}+\sqrt{\alpha_B+C'_{0,1}}}$,
see \cite{JPCS_2023}.
Clearly the coherence length $l_B^{(1)}$ diverges for non-interacting bosons when there is no depletion [$\rho_B^{(1)}=0$].
Illustrative examples and further discussion are given in Sec.~\ref{SEC_APPLICATION}.

Let us proceed and compute
the reduced one-particle density matrix and the density of the bosons in momentum space,
\beqn\label{B_1RDM_DENS_MOM_LP}
& & \tilde\rho^{(1)}_B(q,q') = \frac{1}{2\pi} \int dy dy' e^{-i\left(qy-q'y'\right)} \rho^{(1)}_B(y,y') = \nonumber \\
& & = N_B \left(\frac{1}{\pi\alpha_B}\right)^{\frac{1}{2}}
e^{- \frac{1}{4\alpha_B}\left(q+q'\right)^2}
e^{-\frac{1}{4\left(\alpha_B+C'_{0,1}\right)}\left(q-q'\right)^2}, \nonumber \\
& & \tilde\rho^{(1)}_B(q)
=  N_B \left(\frac{1}{\pi\alpha_B}\right)^{\frac{1}{2}}
e^{- \frac{1}{\alpha_B}q^2}. \
\eeqn
Consequently,
just like (\ref{B_g1_POS_LP}),
the first-order correlation function in momentum space is readily obtained
\beqn\label{B_g1_MOM_LP}
& & \tilde g_B^{(1)}(q,q') = \frac{\tilde\rho^{(1)}_B(q,q')}{\sqrt{\tilde\rho^{(1)}_B(q)\tilde\rho^{(1)}_B(q')}}
= e^{-\frac{-C'_{0,1}}{4\alpha_B\left(\alpha_B+C'_{0,1}\right)}\left(q-q'\right)^2}. \
\eeqn
Analogously to (\ref{l_B_1_POS_LP}), 
we define a dimensionless measure, the first-order coherence length in momentum space,
as the ratio of the off-diagonal decay length of the first-order correlation function (\ref{B_g1_MOM_LP})
and the diagonal decay length of the reduced one-particle density matrix (\ref{B_1RDM_DENS_MOM_LP}), both in momentum space, 
\beqn\label{l_B_1_MOM_LP}
\tilde l_B^{(1)}  
= \sqrt{\frac{\alpha_B+C'_{0,1}}{-C'_{0,1}}} = l_B^{(1)}.
\eeqn
We find that
the first-order coherence lengths in position (\ref{l_B_1_POS_LP}) and momentum  (\ref{l_B_1_MOM_LP}) spaces are equal.

To facilitate the Fourier transform of second-order quantities to momentum space later on,
the reduced two-particle density matrix in position space
is written in a quadratic form using normal coordinates as follows
\beqn\label{B_2RDM_DENS_POS_LP}
& & \rho_B^{(2)}(y_1,y'_1,y_2,y'_2) = 
N_B(N_B-1)\left[\frac{\left(\alpha_B-\beta_B\right)\left(\alpha_B+\beta_B+2C'_{0,2}\right)}{\pi^2}\right]^{\frac{1}{2}}
\times \nonumber \\
& & 
\times e^{-\frac{\alpha_B+\beta_B+2C'_{0,2}}{8}\left(y_1+y_2+y'_1+y'_2\right)^2}
e^{-\frac{\alpha_B-\beta_B}{8}\left(y_1-y_2+y'_1-y'_2\right)^2} \times \nonumber \\
& &
\times
e^{-\frac{\alpha_B+\beta_B}{8}\left(y_1+y_2-y'_1-y'_2\right)^2}
e^{-\frac{\alpha_B-\beta_B}{8}\left(y_1-y_2-y'_1+y'_2\right)^2}, \nonumber \\
& & \rho_B^{(2)}(y_1,y_2) = 
N_B(N_B-1)
\left[\frac{\left(\alpha_B-\beta_B\right)\left(\alpha_B+\beta_B+2C'_{0,2}\right)}{\pi^2}\right]^{\frac{1}{2}}
\times \nonumber \\
& &
\times e^{-\frac{1}{2}\left(\alpha_B+\beta_B+2C'_{0,2}\right)\left(y_1+y_2\right)^2}
e^{-\frac{1}{2}\left(\alpha_B-\beta_B\right)\left(y_1-y_2\right)^2}, \
\eeqn
from which its diagonal, the two-body density in position space, is given \cite{JPCS_2023,arXiv_2023}.

Next, the second-order correlation function in position space reads
\beqn\label{B_g2_POS_LP}
& & g_B^{(2)}(y_1,y_2) = \frac{\rho^{(2)}_B(y_1,y_2)}{\rho^{(1)}_B(y_1)\rho^{(1)}_B(y_2)} = 
\frac{N_B-1}{N_B} \left[\frac{\left(\alpha_B+C'_{0,2}\right)^2}{\left(\alpha_B-\beta_B\right)\left(\alpha_B+\beta_B+2C'_{0,2}\right)}\right]^{\frac{1}{2}} \times \nonumber \\
& & \times e^{-\frac{1}{2}\frac{-\left(\alpha_B-\beta_B\right)\left(\beta_B+C'_{0,2}\right)}{\alpha_B+C'_{0,2}}\left(y_1-y_2\right)^2}
e^{+\frac{1}{2}\frac{-\left(\alpha_B+\beta_B+2C'_{0,2}\right)\left(\beta_B+C'_{0,2}\right)}{\alpha_B+C'_{0,2}}\left(y_1+y_2\right)^2}. \
\eeqn
For $\beta_B+C'_{0,2} < 0$ the off diagonal decays and the diagonal grows,
and for $\beta_B+C'_{0,2} > 0$ vice versa,
also see the applications in Sec.~\ref{SEC_APPLICATION} and \ref{SEC_APPENDIX}.

We can now proceed and define a dimensionless measure, the second-order coherence length in position space,
as the ratio of the off-diagonal decay or growth length of the second-order correlation function (\ref{B_g2_POS_LP})
and the diagonal decay length of the two-particle density (\ref{B_2RDM_DENS_POS_LP}), 
\beqn\label{l_B_2_POS_LP}
& & l_B^{(2)} = \sqrt{\frac{\left(\alpha_B+\beta_B+2C'_{0,2}\right)\left(\alpha_B+C'_{0,2}\right)}
{\left(\alpha_B-\beta_B\right)|\beta_B+C'_{0,2}|}} = \\
& & = \sqrt{\frac{1 + \frac{1}{N_B}\left(\frac{m_BN_B}{M}\frac{\Omega_B}{\omega}+\frac{m_FN_F}{M}\frac{\Omega_B}{\Omega_{BF}}-1\right)}{\left[1 + \frac{2}{N_B}\left(\frac{m_BN_B}{M}\frac{\Omega_B}{\omega}+\frac{m_FN_F}{M}\frac{\Omega_B}{\Omega_{BF}}-1\right)\right]\left|\frac{1}{N_B}\left(\frac{m_BN_B}{M}\frac{\Omega_B}{\omega}+\frac{m_FN_F}{M}\frac{\Omega_B}{\Omega_{BF}}-1\right)\right|}}. \nonumber \
\eeqn
It is instrumental to compare the first-order $l_B^{(1)}$ and second-order $l_B^{(2)}$ coherence lengths,
also see the illustrative examples in Sec.~\ref{SEC_APPLICATION}. 

We now proceed to compute the
the reduced two-particle density matrix and its diagonal, the two-body density of the bosons, in momentum space
which take on the form
\beqn\label{B_2RDM_DENS_MOM_LP}
& & \tilde\rho_B^{(2)}(q_1,q'_1,q_2,q'_2) = \frac{1}{\left(2\pi\right)^2}
\int dy_1 dy_2 dy'_1 dy'_2 e^{-i\left(q_1y_1+q_2y_2-q'_1y'_1-q'_2y'_2\right)} \rho_B^{(2)}(y_1,y'_1,y_2,y'_2) = \nonumber \\
& & = N_B(N_B-1)
\frac{1}{\pi\sqrt{\alpha_B^2-\beta^2_B}}
e^{-\frac{1}{8\left(\alpha_B+\beta_B\right)}\left(q_1+q_2+q'_1+q'_2\right)^2}
e^{-\frac{1}{8\left(\alpha_B-\beta_B\right)}\left(q_1-q_2+q'_1-q'_2\right)^2} \times \nonumber \\
& &
\times
e^{-\frac{1}{8\left(\alpha_B+\beta_B+2C'_{0,2}\right)}\left(q_1+q_2-q'_1-q'_2\right)^2}
e^{-\frac{1}{8\left(\alpha_B-\beta_B\right)}\left(q_1-q_2-q'_1+q'_2\right)^2}, \nonumber \\
& & \tilde\rho_B^{(2)}(q_1,q_2) = 
N_B(N_B-1)
\frac{1}{\pi\sqrt{\alpha_B^2-\beta^2_B}}
e^{-\frac{1}{2\left(\alpha_B+\beta_B\right)}\left(q_1+q_2\right)^2}
e^{-\frac{1}{2\left(\alpha_B-\beta_B\right)}\left(q_1-q_2\right)^2}. \
\eeqn
Then,
the second-order correlation function of the bosons in momentum space can readily be evaluated as
\beqn\label{B_g2_MOM_LP}
& & \tilde g_B^{(2)}(q_1,q_2) = \frac{\tilde\rho^{(2)}_B(q_1,q_2)}{\tilde\rho^{(1)}_B(q_1)\tilde\rho^{(1)}_B(q_2)} =
\nonumber \\
& &
= \frac{N_B-1}{N_B} \sqrt{\frac{\alpha_B^2}{\alpha_B^2-\beta_B^2}}
e^{-\frac{1}{2}\frac{\beta_B}{\alpha_B\left(\alpha_B-\beta_B\right)}\left(q_1-q_2\right)^2}
e^{+\frac{1}{2}\frac{\beta_B}{\alpha_B\left(\alpha_B+\beta_B\right)}\left(q_1+q_2\right)^2}. \
\eeqn
For $\beta_B > 0$ the off diagonal decays and the diagonal grows,
and for $\beta_B < 0$ vice versa,
also see the illustrative examples and \ref{SEC_APPENDIX} below.

Analogously to (\ref{l_B_2_POS_LP}), 
we define a dimensionless measure,
the second-order coherence length in momentum space,
as the ratio of the off-diagonal decay or growth length of the second-order correlation function (\ref{B_g2_MOM_LP})
and the diagonal decay length of the two-particle density (\ref{B_2RDM_DENS_MOM_LP}),
both in momentum space,
\beqn\label{l_B_2_MOM_LP}
& &
\!\!\!\!\!\!\!\!
\tilde l_B^{(2)} = \sqrt{\frac{\alpha_B\left(\alpha_B-\beta_B\right)}
{|\beta_B|\left(\alpha_B+\beta_B\right)}} = \\
& &
\!\!\!\!\!\!\!\!
= \sqrt{\frac{1 + \frac{1}{N_B}\left(\frac{m_BN_B}{M}\frac{\omega}{\Omega_B}+\frac{m_FN_F}{M}\frac{\Omega_{BF}}{\Omega_B}-1\right)}{\left[1 + \frac{2}{N_B}\left(\frac{m_BN_B}{M}\frac{\omega}{\Omega_B}+\frac{m_FN_F}{M}\frac{\Omega_{BF}}{\Omega_B}-1\right)\right]\left|\frac{1}{N_B}\left(\frac{m_BN_B}{M}\frac{\omega}{\Omega_B}+\frac{m_FN_F}{M}\frac{\Omega_{BF}}{\Omega_B}-1\right)\right|}} \ne l_B^{(2)}. \nonumber \
\eeqn
It is found that
the second-order coherence lengths in position (\ref{l_B_2_POS_LP}) and momentum  (\ref{l_B_2_MOM_LP}) spaces are different,
unlike the respective first-order coherence lengths.
The reason is that different parts of the many-particle wavefunction
contribute to the second-order quantities in position and momentum spaces.
Although $l_B^{(1)} \ne \tilde l_B^{(1)}$, 
their respective expressions are `reciprocal' to each other in the sense
$\frac{\Omega_B}{\omega}, \frac{\Omega_B}{\Omega_{BF}} \Leftrightarrow
\frac{\omega}{\Omega_B}, \frac{\Omega_{BF}}{\Omega_B}$, see above
and compare (\ref{l_B_2_POS_LP}) and (\ref{l_B_2_MOM_LP}).
Additionally and on the other hand,
there seems to be no simple relation between the second-order coherence lengths
and the pair depleted fraction $0 \le \rho_B^{(2)} < 1$.
For reference,
the pair depleted fraction is given in closed-form using the above coefficients
and reads 
$\rho_B^{(2)} = \frac{\sqrt{\alpha_B + \beta_B}-\sqrt{\alpha_B+\beta_B+2C'_{0,2}}}{\sqrt{\alpha_B+\beta_B}+\sqrt{\alpha_B+\beta_B+2C'_{0,2}}}$,
see \cite{JPCS_2023}.
Last but not least,
it is deductive to compare the first-order $\tilde l_B^{(1)}$ and second-order $\tilde l_B^{(2)}$
coherence lengths in momentum space as well, also see the illustrative examples in Sec.~\ref{SEC_APPLICATION}. 

\subsection{Position and momentum correlation functions of the fermions}\label{SEC_F}

The reduced one-particle density matrix and the density of the $N_F=2$ fermions in position space read \cite{arXiv_2023}
\beqn\label{F_1RDM_DENS_POS_LP}
& &
\!\!\!\!\!\!\!\!
\rho_F^{(1)}(x,x') = 2 \left(\frac{\alpha_F+C_{1,0}}{\pi}\right)^{\frac{1}{2}} \times \\
& &
\!\!\!\!\!\!\!\!
\times \left\{1-\frac{1}{2}\frac{\alpha_F+C_{1,0}}{\alpha_F-\beta_F}
+ \frac{1}{4}\left[\frac{\left(\alpha_F+C_{1,0}\right)^2}{\alpha_F-\beta_F} \left(x+x'\right)^2
- \left(\alpha_F-\beta_F\right)\left(x-x'\right)^2\right]\right\} \times \nonumber \\
& &
\!\!\!\!\!\!\!\!
\times
e^{-\frac{\alpha_F+C_{1,0}}{4} \left(x+x'\right)^2}
e^{-\frac{\alpha_F}{4}\left(x-x'\right)^2},
\nonumber \\
& &
\!\!\!\!\!\!\!\!
\rho_F^{(1)}(x) =
2 \left(\frac{\alpha_F+C_{1,0}}{\pi}\right)^{\frac{1}{2}} \left[1-\frac{1}{2}\frac{\alpha_F+C_{1,0}}{\alpha_F-\beta_F}
+ \frac{\left(\alpha_F+C_{1,0}\right)^2}{\alpha_F-\beta_F}x^2\right]
e^{-\left(\alpha_F+C_{1,0}\right)x^2}. \nonumber \
\eeqn
Using (\ref{F_1RDM_DENS_POS_LP}),
the first-order correlation function of the fermions in position space is computed as
\beqn\label{F_g1_POS_LP}
& &
\!\!\!\!\!\!\!\!
g_F^{(1)}(x,x') = \frac{\rho^{(1)}_F(x,x')}{\sqrt{\rho^{(1)}_F(x)\rho^{(1)}_F(x')}}
= \nonumber \\
& &
\!\!\!\!\!\!\!\!
= \frac{1-\frac{1}{2}\frac{\alpha_F+C_{1,0}}{\alpha_F-\beta_F}
+ \frac{1}{4}\left[\frac{\left(\alpha_F+C_{1,0}\right)^2}{\alpha_F-\beta_F} \left(x+x'\right)^2
- \left(\alpha_F-\beta_F\right)\left(x-x'\right)^2\right]}{\sqrt{\left[1-\frac{1}{2}\frac{\alpha_F+C_{1,0}}{\alpha_F-\beta_F}
+ \frac{\left(\alpha_F+C_{1,0}\right)^2}{\alpha_F-\beta_F}x^2\right]
\left[1-\frac{1}{2}\frac{\alpha_F+C_{1,0}}{\alpha_F-\beta_F}
+ \frac{\left(\alpha_F+C_{1,0}\right)^2}{\alpha_F-\beta_F}{x'}^2\right]
}}
e^{-\frac{-C_{1,0}}{4}\left(x-x'\right)^2}, \
\eeqn
where $C_{1,0}$ is negative and 
given in closed form as a function of the coefficients (\ref{BF_PSI_CAR_PAR_LP}) \cite{JPA_2017,arXiv_2023},
see for further discussion below.

Similarly to the bosons,
it is instructive to define a dimensionless measure, the first-order coherence length of the fermions in position space,
as the ratio of the off-diagonal decay length of the first-order correlation function (\ref{F_g1_POS_LP})
and the diagonal decay length of the reduced one-particle density matrix (\ref{F_1RDM_DENS_POS_LP}),
giving
\beqn\label{l_F_1_POS_LP}
& &
l_F^{(1)} = \sqrt{\frac{\alpha_F+C_{1,0}}{-C_{1,0}}} = \\
& & 
= \frac{1}{\sqrt{\left[1 + \frac{1}{N_F}\left(\frac{m_FN_F}{M}\frac{\omega}{\Omega_F}+\frac{m_BN_B}{M}\frac{\Omega_{BF}}{\Omega_F}-1\right)\right]\left[1 + \frac{1}{N_F}\left(\frac{m_FN_F}{M}\frac{\Omega_F}{\omega}+\frac{m_BN_B}{M}\frac{\Omega_F}{\Omega_{BF}}-1\right)\right]-1}}. \nonumber \
\eeqn
Formally, it is possible to re-express the first-order
coherence length (\ref{l_F_1_POS_LP}) using the quantity 
$\bar \rho_F^{(1)} = \frac{\sqrt{\alpha_F}-\sqrt{\alpha_F+C_{1,0}}}{\sqrt{\alpha_F}+\sqrt{\alpha_F+C_{1,0}}}$,
which satisfies $0 \le \bar\rho_F^{(1)} < 1$ and
which would have been the depleted fraction if the impurity were bosonic, see \cite{JPCS_2023}.
The final result is $l_F^{(1)}=\left[\left(\frac{1+\bar\rho_F^{(1)}}{1-\bar\rho_F^{(1)}}\right)^2-1\right]^{-\frac{1}{2}}$,
compare to (\ref{l_B_1_POS_LP}).
Obviously, the first-order coherence length $l_F^{(1)}$ diverges for non-interacting fermions.

The reduced one-particle density matrix and the density of the fermions in momentum space
are readily evaluated from the reduced one-particle density matrix in position space,
\beqn\label{F_1RDM_DENS_MOM_LP}
& &
\!\!\!\!\!\!\!\!
\tilde\rho_F^{(1)}(k,k') = \frac{1}{2\pi} \int dx dx' e^{-i\left(kx-k'x'\right)} \rho^{(1)}_F(x,x') = \nonumber \\
& &
\!\!\!\!\!\!\!\!
= 2 \left(\frac{1}{\pi\alpha_F}\right)^{\frac{1}{2}}
\left\{1-\frac{1}{2}\frac{\alpha_F-\beta_F}{\alpha_F}
+ \frac{1}{4}\left[\frac{\alpha_F-\beta_F}{\alpha_F^2} \left(k+k'\right)^2
- \frac{1}{\alpha_F-\beta_F}\left(k-k'\right)^2\right]\right\} \times \nonumber \\
& &
\!\!\!\!\!\!\!\!
\times
e^{- \frac{1}{4\alpha_F}\left(k+k'\right)^2}
e^{-\frac{1}{4\left(\alpha_F+C_{1,0}\right)}\left(k-k'\right)^2},
\nonumber \\
& &
\!\!\!\!\!\!\!\!
\tilde\rho_F^{(1)}(k) =
2 \left(\frac{1}{\pi\alpha_F}\right)^{\frac{1}{2}} \left[1-\frac{1}{2}\frac{\alpha_F-\beta_F}{\alpha_F}
+ \frac{\alpha_F-\beta_F}{\alpha_F^2}k^2\right]
e^{-\frac{1}{\alpha_F}k^2}. \
\eeqn
Afterwards,
the first-order correlation function in momentum space is given by
\beqn\label{F_g1_MOM_LP}
& &
\!\!\!\!\!\!\!\!
\tilde g_F^{(1)}(k,k') = \frac{\tilde\rho^{(1)}_F(k,k')}{\sqrt{\tilde\rho^{(1)}_F(k)\tilde\rho^{(1)}_F(k')}}
= \nonumber \\
& & 
\!\!\!\!\!\!\!\!
= \frac{1-\frac{1}{2}\frac{\alpha_F-\beta_F}{\alpha_F}
+ \frac{1}{4}\left[\frac{\alpha_F-\beta_F}{\alpha_F^2} \left(k+k'\right)^2
- \frac{1}{\alpha_F-\beta_F}\left(k-k'\right)^2\right]}{\sqrt{\left[1-\frac{1}{2}\frac{\alpha_F-\beta_F}{\alpha_F}
+ \frac{\alpha_F-\beta_F}{\alpha_F^2}k^2\right]
\left[1-\frac{1}{2}\frac{\alpha_F-\beta_F}{\alpha_F}
+ \frac{\alpha_F-\beta_F}{\alpha_F^2}{k'}^2\right]
}}
e^{-\frac{-C_{1,0}}{4\alpha_F\left(\alpha_F+C_{1,0}\right)}\left(k-k'\right)^2}. \
\eeqn
Just like (\ref{l_B_1_MOM_LP}), 
we define a dimensionless measure,
the first-order coherence length of the fermions in momentum space,
as the ratio of the off-diagonal decay length of the first-order correlation function (\ref{F_g1_MOM_LP})
and the diagonal decay length of the reduced one-particle density matrix (\ref{F_1RDM_DENS_MOM_LP}), both in momentum space,
\beqn\label{l_F_1_MOM_LP}
\tilde l_F^{(1)} = \sqrt{\frac{\alpha_F+C_{1,0}}{-C_{1,0}}} = l_F^{(1)}.
\eeqn
It is found that
the first-order coherence lengths in position (\ref{l_F_1_POS_LP}) and momentum  (\ref{l_F_1_MOM_LP}) spaces are equal.

To ease the Fourier transform of second-order quantities to momentum space below,
the reduced two-particle density matrix in position space,
including the term $\left(x_1-x_2\right)\left(x'_1-x'_2\right)$,
is expressed in a quadratic form using normal coordinates.
The final result is given by
\beqn\label{F_2RDM_DENS_POS_LP}
& & \rho_F^{(2)}(x_1,x'_1,x_2,x'_2) = 
2 \left[\frac{\left(\alpha_F-\beta_F\right)^3\left(\alpha_F+\beta_F+2C_{2,0}\right)}{\pi^2}\right]^{\frac{1}{2}}
\times \nonumber \\
& & 
\times \left(x_1-x_2\right)\left(x'_1-x'_2\right)
e^{-\frac{\alpha_F+\beta_F+2C_{2,0}}{8}\left(x_1+x_2+x'_1+x'_2\right)^2}
e^{-\frac{\alpha_F-\beta_F}{8}\left(x_1-x_2+x'_1-x'_2\right)^2} \times \nonumber \\
& &
\times
e^{-\frac{\alpha_F+\beta_F}{8}\left(x_1+x_2-x'_1-x'_2\right)^2}
e^{-\frac{\alpha_F-\beta_F}{8}\left(x_1-x_2-x'_1+x'_2\right)^2}, \nonumber \\
& & \rho_F^{(2)}(x_1,x_2) =
2 \left[\frac{\left(\alpha_F-\beta_F\right)^3\left(\alpha_F+\beta_F+2C_{2,0}\right)}{\pi^2}\right]^{\frac{1}{2}}
\times \nonumber \\
& &
\times \left(x_1-x_2\right)^2
e^{-\frac{1}{2}\left(\alpha_F+\beta_F+2C_{2,0}\right)\left(x_1+x_2\right)^2}
e^{-\frac{1}{2}\left(\alpha_F-\beta_F\right) \left(x_1-x_2\right)^2}, \
\eeqn
from which the two-body density in position space is prescribed.

The second-order correlation function in position space is now evaluated as 
\beqn\label{F_g2_POS_LP}
& & g_F^{(2)}(x_1,x_2) = \frac{\rho^{(2)}_F(x_1,x_2)}{\rho^{(1)}_F(x_1)\rho^{(1)}_F(x_2)} =
\frac{1}{2}
\left[\frac{\left(\alpha_F+C_{2,0}\right)^2}{\left(\alpha_F-\beta_F\right)^3\left(\alpha_F+\beta_F+2C_{2,0}\right)}\right]^{\frac{1}{2}} \times \nonumber \\
& & \times \frac{\left(x_1-x_2\right)^2}{\left[\frac{1}{2\left(\alpha_F+C_{2,0}\right)} + \left(1+\frac{\beta_F+C_{2,0}}{\alpha_F+C_{2,0}}\right)^2x_1^2\right]\left[\frac{1}{2\left(\alpha_F+C_{2,0}\right)} + \left(1+\frac{\beta_F+C_{2,0}}{\alpha_F+C_{2,0}}\right)^2x_2^2\right]}
\times \nonumber \\
& & \times e^{-\frac{1}{2}\frac{-\left(\alpha_F-\beta_F\right)\left(\beta_F+C_{2,0}\right)}{\alpha_F+C_{2,0}}\left(x_1-x_2\right)^2}
e^{+\frac{1}{2}\frac{-\left(\alpha_F+\beta_F+2C_{2,0}\right)\left(\beta_F+C_{2,0}\right)}{\alpha_F+C_{2,0}}\left(x_1+x_2\right)^2}. \
\eeqn
For $\beta_F+C_{2,0} < 0$ the off diagonal decays and the diagonal grows
whereas the opposite occurs for 
$\beta_F+C_{2,0} > 0$,
also see Sec.~\ref{SEC_APPLICATION} and \ref{SEC_APPENDIX} below.

We define and compute a dimensionless measure,
the second-order coherence length of the fermions in position space,
as the ratio of the off-diagonal decay or growth length of the second-order correlation function (\ref{F_g2_POS_LP})
and the diagonal decay length of the two-particle density (\ref{F_2RDM_DENS_POS_LP}), 
\beqn\label{l_F_2_POS_LP}
& &
l_F^{(2)} = \sqrt{\frac{\left(\alpha_F+\beta_F+2C_{2,0}\right)\left(\alpha_F+C_{2,0}\right)}
{\left(\alpha_F-\beta_F\right)|\beta_F+C_{2,0}|}} = \\
& &
= \sqrt{\frac{1 + \frac{1}{N_F}\left(\frac{m_FN_F}{M}\frac{\Omega_F}{\omega}+\frac{m_BN_B}{M}\frac{\Omega_F}{\Omega_{BF}}-1\right)}{\left[1 + \frac{2}{N_F}\left(\frac{m_FN_F}{M}\frac{\Omega_F}{\omega}+\frac{m_BN_B}{M}\frac{\Omega_F}{\Omega_{BF}}-1\right)\right]\left|\frac{1}{N_F}\left(\frac{m_FN_F}{M}\frac{\Omega_F}{\omega}+\frac{m_BN_B}{M}\frac{\Omega_F}{\Omega_{BF}}-1\right)\right|}}. \nonumber \
\eeqn
It would be instructive
to contrast the first-order $l_F^{(1)}$ and second-order $l_F^{(2)}$ coherence lengths,
which is done in the illustrative examples of Sec.~\ref{SEC_APPLICATION}. 

To continue,
the reduced two-particle density matrix and two-body density of the fermions in momentum space are calculated,
\beqn\label{F_2RDM_DENS_MOM_LP}
& &
\!\!\!\!\!\!\!\!\!\!\!\!\!\!\!\!
\tilde\rho_F^{(2)}(k_1,k'_1,k_2,k'_2) = \frac{1}{\left(2\pi\right)^2}
\int dx_1 dx_2 dx'_1 dx'_2 e^{-i\left(k_1x_1+k_2x_2-k'_1x'_1-k'_2x'_2\right)} \rho_F^{(2)}(x_1,x'_1,x_2,x'_2) = \nonumber \\
& &
\!\!\!\!\!\!\!\!\!\!\!\!\!\!\!\!
= 2 \frac{1}{\pi\sqrt{\left(\alpha_F^2-\beta^2_F\right)\left(\alpha_F-\beta_F\right)^2}}
\left(k_1-k_2\right)\left(k'_1-k'_2\right)
e^{-\frac{1}{8\left(\alpha_F+\beta_F\right)}\left(k_1+k_2+k'_1+k'_2\right)^2} \times \\
& &
\!\!\!\!\!\!\!\!\!\!\!\!\!\!\!\!
\times
e^{-\frac{1}{8\left(\alpha_F-\beta_F\right)}\left(k_1-k_2+k'_1-k'_2\right)^2} 
e^{-\frac{1}{8\left(\alpha_F+\beta_F+2C_{2,0}\right)}\left(k_1+k_2-k'_1-k'_2\right)^2}
e^{-\frac{1}{8\left(\alpha_F-\beta_F\right)}\left(k_1-k_2-k'_1+k'_2\right)^2}, \nonumber \\
& &
\!\!\!\!\!\!\!\!\!\!\!\!\!\!\!\!
\tilde\rho_F^{(2)}(k_1,k_2) = 
2 \frac{1}{\pi\sqrt{\left(\alpha_F^2-\beta^2_F\right)\left(\alpha_F-\beta_F\right)^2}}
\left(k_1-k_2\right)^2
e^{-\frac{1}{2\left(\alpha_F+\beta_F\right)}\left(k_1+k_2\right)^2}
e^{-\frac{1}{2\left(\alpha_F-\beta_F\right)}\left(k_1-k_2\right)^2}. \nonumber \
\eeqn
Thereafter,
the second-order correlation function in momentum space is evaluated and reads
\beqn\label{F_g2_MOM_LP}
& & \tilde g_F^{(2)}(k_1,k_2) = \frac{\tilde\rho^{(2)}_F(k_1,k_2)}{\tilde\rho^{(1)}_F(k_1)\tilde\rho^{(1)}_F(k_2)} =
\frac{1}{2} \sqrt{\frac{\alpha_F^2}{\left(\alpha_F^2-\beta^2_F\right)\left(\alpha_F-\beta_F\right)^2}} \times
\nonumber \\
& &
\times \frac{\left(k_1-k_2\right)^2}{\left[1-\frac{1}{2}\frac{\alpha_F-\beta_F}{\alpha_F} + \frac{\alpha_F-\beta_F}{\alpha_F^2}k_1^2\right]
\left[1-\frac{1}{2}\frac{\alpha_F-\beta_F}{\alpha_F} + \frac{\alpha_F-\beta_F}{\alpha_F^2}k_2^2\right]} \times
\nonumber \\
& &
\times e^{-\frac{1}{2}\frac{\beta_F}{\alpha_F\left(\alpha_F-\beta_F\right)}\left(k_1-k_2\right)^2}
e^{+\frac{1}{2}\frac{\beta_F}{\alpha_F\left(\alpha_F+\beta_F\right)}\left(k_1+k_2\right)^2}. \
\eeqn
For $\beta_F > 0$ the off diagonal decays and the diagonal grows,
and for $\beta_F < 0$ the other way around,
also see the applications below.

Finally,
a dimensionless measure, the second-order coherence length of the fermions in momentum space,
is defined as the ratio of the off-diagonal decay or growth length of the second-order momentum correlation function (\ref{F_g2_MOM_LP})
and the diagonal decay length of the two-particle momentum density (\ref{F_2RDM_DENS_MOM_LP})
and found to be
\beqn\label{l_F_2_MOM_LP}
& &
\!\!\!\!\!\!\!\!
\tilde l_F^{(2)} = \sqrt{\frac{\alpha_F\left(\alpha_F-\beta_F\right)}
{|\beta_F|\left(\alpha_F+\beta_F\right)}} = \\
& &
\!\!\!\!\!\!\!\!
= \sqrt{\frac{1 + \frac{1}{N_F}\left(\frac{m_FN_F}{M}\frac{\omega}{\Omega_F}+\frac{m_BN_B}{M}\frac{\Omega_{BF}}{\Omega_F}-1\right)}{\left[1 + \frac{2}{N_F}\left(\frac{m_FN_F}{M}\frac{\omega}{\Omega_F}+\frac{m_BN_B}{M}\frac{\Omega_{BF}}{\Omega_F}-1\right)\right]\left|\frac{1}{N_F}\left(\frac{m_FN_F}{M}\frac{\omega}{\Omega_F}+\frac{m_BN_B}{M}\frac{\Omega_{BF}}{\Omega_F}-1\right)\right|}} \ne l_F^{(2)}. \nonumber \
\eeqn
This implies that
that the second-order coherence lengths in position (\ref{l_F_2_POS_LP}) and momentum (\ref{l_F_2_MOM_LP}) spaces are different,
just as for the bosons.
Although $l_F^{(1)} \ne \tilde l_F^{(1)}$, 
their respective expressions are `reciprocal' to each other
in the manner $\frac{\Omega_F}{\omega}, \frac{\Omega_F}{\Omega_{BF}} \Leftrightarrow
\frac{\omega}{\Omega_F}, \frac{\Omega_{BF}}{\Omega_F}$, see above
and compare (\ref{l_F_2_POS_LP}) and (\ref{l_F_2_MOM_LP}).
Finally and unlike the above analysis for the first-order coherence lengths,
there is probably no simple relation between the second-order coherence lengths
and between what would have been the pair depleted fraction $0 \le \frac{\sqrt{\alpha_F + \beta_F}-\sqrt{\alpha_F+\beta_F+2C_{2,0}}}{\sqrt{\alpha_F+\beta_F}+\sqrt{\alpha_F+\beta_F+2C_{2,0}}} < 1$, see \cite{JPCS_2023},
if the impurity were made of bosons.
It would hence be 
rewarding to compare the first-order $\tilde l_F^{(1)}$ and second-order $\tilde l_F^{(2)}$ coherence lengths in momentum space as well, also examine the application section below.

\section{Application}\label{SEC_APPLICATION}

\begin{figure}[!]
\begin{center}
\hglue -1.2 truecm
\includegraphics[width=0.42\columnwidth,angle=-90]{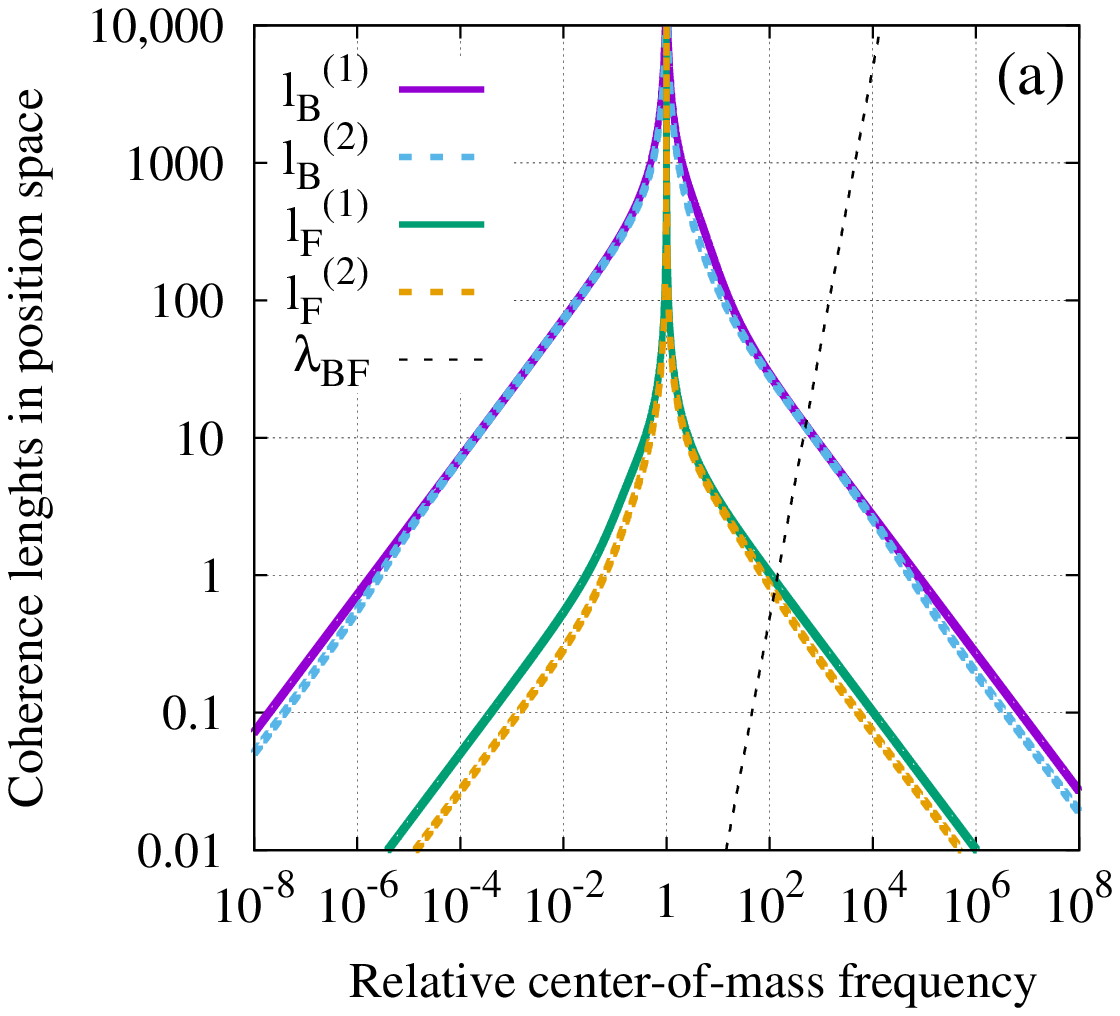}
\hglue -2.12 truecm
\includegraphics[width=0.42\columnwidth,angle=-90]{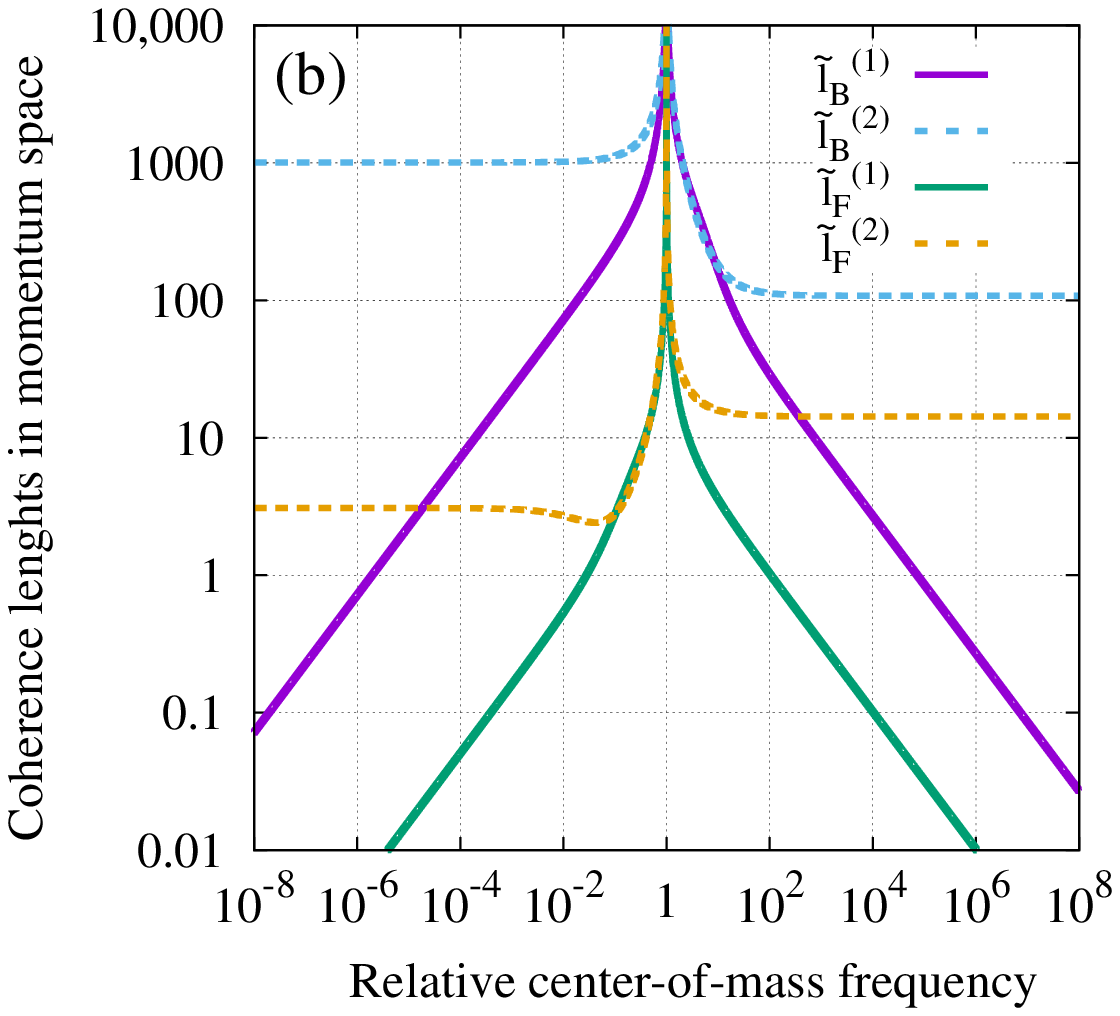}
\end{center}
\vglue 0.5 truecm
\caption{First-order and second-order dimensionless coherence lengths
of the impurity and the condensate.
The number of fermions is $N_F=2$ and the number of bosons $N_B=10,000$.
The mass of a fermion is $m_F=1.0$ and that of a boson $m_B=0.01$.
Panel (a) depicts the coherence lengths in position space and panel (b) in momentum space.
The results are plotted against the relative center-of-mass frequency
$\Omega_{BF}=\sqrt{\omega^2 + 2\left(\frac{N_B}{m_F}+\frac{N_F}{m_B}\right)\lambda_{BF}}$ [Eq.~(\ref{BF_FREQ_LP})]
which satisfies $1 < \Omega_{BF}$ for attractive and $0^+ < \Omega_{BF} < 1$ for repulsive interspecies interactions.
For reference, in panel (a) the interspecies interaction strength
$\lambda_{BF}=\frac{1}{2}\frac{m_Fm_B}{M}\left(\Omega^2_{BF}-\omega^2\right)$ is plotted.
The frequency of the trap is $\omega=1.0$.
There are no intraspecies interactions, i.e., $\lambda_F=0$ and $\lambda_B=0$.
Correlations build up only due to impurity-condensate interaction.
The first-order coherence lengths in position and momentum spaces are equal
whereas the second-order quantities can differ considerably.
See the text for more details.
The quantities shown are dimensionless.}
\label{F1_LP}
\end{figure}

We present illustrative examples.
The results are shown in Fig.~\ref{F1_LP}.
Consider an impurity made of $N_F=2$ fermions
in a condensate made of $N_B=10,000$ bosons.
The mass of a fermion is $m_F=1.0$ and that of a boson $m_B=0.01$.
The frequency of the trap is $\omega=1.0$.
We explore the case where there are no intraspecies interactions,
namely, $\lambda_F=0$ and $\lambda_B=0$.
The correlation effects are generated only by the interspecies interaction which can assume the values
$-\frac{M_{BF}\omega^2}{2} < \lambda_{BF}$.
We investigate the above-derived eight coherence lengths,
$l_B^{(1)}, l_B^{(2)}, l_F^{(1)}, l_F^{(2)}$ in position space in Fig.~\ref{F1_LP}a and 
$\tilde l_B^{(1)}, \tilde l_B^{(2)}, \tilde l_F^{(1)}, \tilde l_F^{(2)}$ in momentum space in Fig.~\ref{F1_LP}b.
It is convenient to plot quantities as a function of the relative center-of-mass frequency $\Omega_{BF}=\sqrt{\omega^2 + 2\left(\frac{N_B}{m_F}+\frac{N_F}{m_B}\right)\lambda_{BF}}$, see Eq.~(\ref{BF_FREQ_LP}),
which in the attractive sector is not bound from above and satisfies $1 < \Omega_{BF}$ and in the repulsive sector obeys $0^+ < \Omega_{BF} < 1$.
For completeness, the interspecies interaction strength
$\lambda_{BF}=\frac{1}{2}M_{BF}\left(\Omega^2_{BF}-\omega^2\right)$ is also plotted in Fig.~\ref{F1_LP}
(in the repulsive sector for the above parameters, 
$|\lambda_{BF}| < \frac{M_{BF}\omega^2}{2} \approx 4.90 \cdot 10^{-5}$ and is not visible).
Obviously, for $\Omega_{BF}=1$ the mixture is non-interacting ($\lambda_{BF}=0$),
and correspondingly all coherence lengths diverge per definition.

As the impurity-condensate interaction sets in,
whether repulsive or attractive,
all coherence lengths become finite, see Fig.~\ref{F1_LP}.
A diverse behavior is found.
In position space, Fig.~\ref{F1_LP}a, 
the impurity and condensate coherence lengths decay to zero with
increasing interspecies interaction, repulsive or attractive,
and the second-order coherence lengths are nearly the same
as the first-order ones.
In momentum space, Fig.~\ref{F1_LP}b,
as shown above in (\ref{l_B_1_MOM_LP}) and (\ref{l_F_1_MOM_LP}),
the first-order coherence lengths are equal to their position-space counterparts and thus decay to zero.
On the other hand,
both second-order coherence lengths $\tilde l_B^{(2)}$ and $\tilde l_F^{(2)}$
saturate to (different) finite values for strong interspecies repulsion or attraction.
Analysis of the closed-form expressions for $\tilde l_B^{(2)}$ and $\tilde l_F^{(2)}$,
Eqs.~(\ref{l_B_2_MOM_LP}) and (\ref{l_F_2_MOM_LP}), 
straightforwardly show why the limiting values of the second-order momentum-space coherence lengths are finite,
whereas the other six coherence lengths decay to zero in these limits.
We finish with a question and the beginning of its answer.
Can one have different behaviors of the coherence lengths than the above?
The answer is positive.
Two other scenarios are analyzed in the \ref{SEC_APPENDIX},
demonstrating and emphasizing further the richness
of the exactly-solvable harmonic-interaction many-body
model for Bose-Fermi mixtures \cite{arXiv_2023}.

\section{Summary}\label{SEC_SUM}

A fermionic impurity embedded in a Bose-Einstein condensate is treated analytically,
at the many-body level of theory,
within the exactly-solvable harmonic-interaction model
for trapped Bose-Fermi mixtures.
In the present work,
we have focused on the construction and analysis
of correlation functions, both in position and momentum spaces,
for $N_F=2$ fermions and a general number $N_B$ of bosons.
For the fermionic impurity as well as for the condensate,
we have prescribed the exact first-order and second-order correlation
functions as functions of the masses $m_F, m_B$ and intraspecies $\lambda_F, \lambda_B$ and interspecies $\lambda_{BF}$
interaction strengths.
Furthermore, we have studied properties of the various coherence lengths,
defined as the ratio between the off-diagonal decay or growth length of a correlation function and the diagonal decay
length of the respective (reduced) density.
It has been found that the first-order coherence lengths in position and momentum spaces are equal,
whereas the second-order quantities can differ significantly.
Illustrative examples were presented and discussed.
In particular,
when the sole interaction is between the impurity and the condensate,
all coherence lengths decay to zero 
for strong attraction or repulsion,
except for the second-order coherence lengths in momentum space which saturate at finite values.
As an obvious yet challenging outlook,
it would be interesting
to extend the Bose-Fermi harmonic-interaction model and its applications
to time-dependent scenarios.
 
\ack

This research was supported by the Israel Science Foundation (Grant No. 1516/19).

\appendix

\section{Limiting cases of the bosons' and fermions' first-order and second-order dimensionless coherence lengths
in position and momentum spaces}\label{SEC_APPENDIX}

Consider, first, the case of no interaction between the impurity and the condensate,
i.e., $\Omega_{BF}=\omega$.
We analyze the bosons but the results hold for the fermions as well.
Then,
\beqn\label{l_B_1_LP_Single}
& &
l_B^{(1)} = \tilde l_B^{(1)} =
\frac{1}{\sqrt{\left[1 + \frac{1}{N_B}\left(\frac{\omega}{\Omega_B}-1\right)\right]\left[1 + \frac{1}{N_B}\left(\frac{\Omega_B}{\omega}-1\right)\right]-1}} \
\eeqn
and
\beqn\label{l_B_2_LP_Single}
& &
\!\!\!\!\!\!\!\!\!\!\!\!\!\!\!\!
l_B^{(2)} = \sqrt{\frac{1 + \frac{1}{N_B}\left(\frac{\Omega_B}{\omega}-1\right)}{\left[1 + \frac{2}{N_B}\left(\frac{\Omega_B}{\omega}-1\right)\right]\left|\frac{1}{N_B}\left(\frac{\Omega_B}{\omega}-1\right)\right|}}, \quad
\tilde l_B^{(2)} = \sqrt{\frac{1 + \frac{1}{N_B}\left(\frac{\omega}{\Omega_B}-1\right)}{\left[1 + \frac{2}{N_B}\left(\frac{\omega}{\Omega_B}-1\right)\right]\left|\frac{1}{N_B}\left(\frac{\omega}{\Omega_B}-1\right)\right|}}. \nonumber \\ \
\eeqn
The limits $\lim_{\Omega_B \to \infty}$ and $\lim_{\Omega_B \to 0^+}$
of the coherence lengths (\ref{l_B_1_LP_Single}) and (\ref{l_B_2_LP_Single}) are all equal to zero
expect for
\beqn\label{l_B_1_2_LP_Single_LIM}
\lim_{\Omega_B \to 0^+} l_B^{(2)} =
\lim_{\Omega_B \to \infty} \tilde l_B^{(2)} = \sqrt{\frac{1-\frac{1}{N_B}}{\left(1-\frac{2}{N_B}\right)\frac{1}{N_B}}}.
\
\eeqn
The meaning of (\ref{l_B_1_2_LP_Single_LIM}) is that the second-order position and momentum
coherence lengths are always finite (for a finite number of particles) at the limit
of strong attractive and repulsive interactions, respectively,
expect for a single pair of identical particles, $N_B=2$.
Then, the coherence length diverges because the diagonal of the two-particle density
is nothing but the pairs' center-of-mass wavefunction (density) which does not depend on the interaction. 

Next, consider the regime of parameters in which the interspecies interaction is `balanced' by
the intraspecies interactions,
namely
$\lambda_F=-\frac{N_B}{N_F}\lambda_{BF}$ and
$\lambda_B=-\frac{N_F}{N_B}\lambda_{BF}$.
Consequently, $\Omega_F=\Omega_B=\omega$.
Furthermore,
the energy (\ref{BF_Energy_LP}) depends explicitly on the interspecies
interaction only though the relative center-of-mass degree-of-freedom,
\beqn\label{BF_Energy_LP_MARG}
& &
E = \frac{1}{2} \left[\left(N_F^2+N_B-1\right)\omega + 
\sqrt{\omega^2 + 2\left(\frac{N_B}{m_F}+\frac{N_F}{m_B}\right)\lambda_{BF}}\right].
\eeqn
Thus, we may call this regime of `balanced' interactions the regime of `marginal interaction',
since only one out of $N_F+N_B$ degrees-of-freedom is explicitly interaction dependent.
Now,
\beqn\label{l_B_1_LP_MARG}
& &
l_B^{(1)} = \tilde l_B^{(1)} =
\frac{1}{\sqrt{\left[1 + \frac{1}{N_B}\frac{m_FN_F}{M}\left(\frac{\Omega_{BF}}{\omega}-1\right)\right]\left[1 + \frac{1}{N_B}\frac{m_FN_F}{M}\left(\frac{\omega}{\Omega_{BF}}-1\right)\right]-1}} \
\eeqn
and
\beqn\label{l_B_2_LP_MARG}
& &
l_B^{(2)} = \sqrt{\frac{1 + \frac{1}{N_B}\frac{m_FN_F}{M}\left(\frac{\omega}{\Omega_{BF}}-1\right)}{\left[1 + \frac{2}{N_B}\frac{m_FN_F}{M}\left(\frac{\omega}{\Omega_{BF}}-1\right)\right]\left|\frac{1}{N_B}\frac{m_FN_F}{M}\left(\frac{\omega}{\Omega_{BF}}-1\right)\right|}}, \nonumber \\
& &
\tilde l_B^{(2)} = \sqrt{\frac{1 + \frac{1}{N_B}\frac{m_FN_F}{M}\left(\frac{\Omega_{BF}}{\omega}-1\right)}{\left[1 + \frac{2}{N_B}\frac{m_FN_F}{M}\left(\frac{\Omega_{BF}}{\omega}-1\right)\right]\left|\frac{1}{N_B}\frac{m_FN_F}{M}\left(\frac{\Omega_{BF}}{\omega}-1\right)\right|}}. \
\eeqn
The limits $\lim_{\Omega_{BF} \to \infty}$ and $\lim_{\Omega_{BF} \to 0^+}$
of the coherence lengths (\ref{l_B_1_LP_MARG}) and (\ref{l_B_2_LP_MARG}) are all equal to zero
expect for
\beqn\label{l_B_1_2_LP_MARG_LIM}
\lim_{\Omega_{BF} \to \infty} l_B^{(2)} =
\lim_{\Omega_{BF} \to 0^+} \tilde l_B^{(2)} = \sqrt{\frac{1-\frac{1}{N_B}\frac{m_FN_F}{M}}{\left(1-\frac{2}{N_B}\frac{m_FN_F}{M}\right)\frac{1}{N_B}\frac{m_FN_F}{M}}},
\
\eeqn
which are finite for any number of particles $N_B$ including a pair, compare to (\ref{l_B_1_2_LP_Single_LIM}).
The interspecies interaction dresses even a single pair
of the other species,
making the pair's center-of-mass degree-of-freedom interaction dependent.
Consequently, the respective second-order coherence lengths become
finite at the limit of strong attractive and repulsive interactions,
see (\ref{l_B_1_2_LP_MARG_LIM}).

Finally, equivalent limiting relations as the above,
Eqs.~(\ref{l_B_1_LP_Single}-\ref{l_B_1_2_LP_Single_LIM}) and (\ref{l_B_1_LP_MARG}-\ref{l_B_1_2_LP_MARG_LIM}),
hold for the coherence lengths of the fermions,
$l_F^{(1)} = \tilde l_F^{(1)}$, $l_F^{(2)}$, and $\tilde l_F^{(2)}$,
with the obvious interchanges $N_B \Leftrightarrow N_F$,
$\Omega_B \Leftrightarrow \Omega_F$,
and
$m_F \Leftrightarrow m_B$.


\section*{References}
\begin{thebibliography}{99}

\bibitem{bf1} Bijlsma M J, Heringa B A and Stoof H T C 2000
                   {\it Phys. Rev. A} {\bf 61} 053601

\bibitem{bf2} Lewenstein M, Santos L, Baranov M A and Fehrmann H 2004
                    {\it Phys. Rev. Lett.} {\bf 92} 050401

\bibitem{bf3} Frahm H and Palacios G 2005
                    {\it Phys. Rev. A} {\bf 72} 061604(R)

\bibitem{bf4} Imambekov A and Demler E 2006
                    {\it Phys. Rev. A} {\bf 73} 021602(R)

\bibitem{bf5} G\"unter K, St\"oferle T, Moritz H, K\"ohl M and Esslinger T 2006
                    {\it Phys. Rev. Lett.} {\bf 96} 180402

\bibitem{bf6} Ospelkaus S, Ospelkaus C, Wille O, Succo M, Ernst P, Sengstock K and Bongs K 2006
                    {\it Phys. Rev. Lett.} {\bf 96} 180403

\bibitem{bf7} Zaccanti M, D'Errico C, Ferlaino F, Roati G, Inguscio M and Modugno G 2006
                    {\it Phys. Rev. A} {\bf 74} 041605(R)

\bibitem{bf8} Alon O E, Streltsov A I and Cederbaum L S 2007
                    {\it Phys. Rev. A} {\bf 76} 062501

\bibitem{bf9} Girardeau M D and Minguzzi A 2007
                    {\it Phys. Rev. Lett.} {\bf 99} 230402

\bibitem{bf10} Pollet L, Kollath C, Schollw\"ock U and Troyer M 2008
                    {\it Phys. Rev. A} {\bf 77} 023608

\bibitem{bf11} Adhikari S K and Salasnich L 2008
                     {\it Phys. Rev. A} {\bf 78} 043616

\bibitem{bf12} Fang B, Vignolo P, Gattobigio M, Miniatura C and Minguzzi A 2011
                      {\it Phys. Rev. A} {\bf 84} 023626

\bibitem{bf13} Heinze J, G\"otze S, Krauser J S, Hundt B, Fl\"aschner N, L\"uhmann D-S, Becker C and Sengstock K 2011
                      {\it Phys. Rev. Lett.} {\bf 107} 135303

\bibitem{bf14} Wang H, Hao Y and Zhang Y 2012
                      {\it Phys. Rev. A} {\bf 85} 053630

\bibitem{bf15} Yamamoto A and Hatsuda T 2012
                      {\it Phys. Rev. A} {\bf 86} 043627

\bibitem{bf16} Bloom R S, Hu M-G, Cumby T D and Jin D S 2013
                      {\it Phys. Rev. Lett.} {\bf 111} 105301

\bibitem{bf17} Wen L and Li J 2014
                      {\it Phys. Rev. A} {\bf 90} 053621

\bibitem{bf18} Vaidya V D, Tiamsuphat J, Rolston S L and Porto J V 2015
                     {\it Phys. Rev. A} {\bf 92} 043604

\bibitem{bf19} Bradlyn B and Andrey G 2016
                      {\it Phys. Rev. A} {\bf 93} 033642

\bibitem{bf20} Cao L, Bolsinger V, Mistakidis S I, Koutentakis G M, Kr\"onke S, Schurer J M and Schmelcher P 2017
                     {\it J. Chem. Phys.} {\bf 147} 044106

\bibitem{bf21} DeSalvo B J, Patel K, Johansen J and Chin C 2017
                      {\it Phys. Rev. Lett.} {\bf 119} 233401

\bibitem{bf22} Chen J, Schurer J M and Schmelcher P 2018
                     {\it Phys. Rev. Lett.} {\bf 121} 043401

\bibitem{bf23} Trautmann A, Ilzh\"ofer P, Durastante G, Politi C, Sohmen M, Mark M J and Ferlaino F 2018
                      {\it Phys. Rev. Lett.} {\bf 121} 213601

\bibitem{bf24} Kinnunen J J, Wu Z and Bruun G M 2018
                      {\it Phys. Rev. Lett.} {\bf 121} 253402

\bibitem{bf25} Gautam S and Adhikari S K 2019
                     {\it Phys. Rev. A} {\bf 100} 023626

\bibitem{bf26} Pasek M and Orso G 2019
                     {\it Phys. Rev. B} {\bf 100} 245419

\bibitem{bf27} Sowi\'nski T and Garc\'ia-March M \'A 2019
                     {\it Rep. Prog. Phys.} {\bf 82} 104401

\bibitem{bf28} Ye Z-X, Xie L-Y, Guo Z, Ma X-B, Wang G-R, You L and Tey M K 2020
                      {\it Phys. Rev. A} {\bf 102} 033307
 
\bibitem{bf29} Jee K Y and Mueller E 2021
                      {\it Phys. Rev. A} {\bf 103} 033307

\bibitem{bf30} Milczewski J von, Rose F and Schmidt R 2022
                     {\it Phys. Rev. A} {\bf 105} 013317

\bibitem{bf31} P\^a\c{t}u O I 2022
                      {\it Phys. Rev. A} {\bf 105} 063309

\bibitem{bf32} Patel K, Cai G, Ando H and Chin C 2023
                      {\it Phys. Rev. Lett.} {\bf 131} 083003

\bibitem{him1} Pruski S, Ma\'ckowiak J and Missuno O 1972
                     {\it Rep. Math. Phys.} {\bf 3} 227

\bibitem{him2} Robinson P D 1977
                      {\it J. Chem. Phys.} {\bf 66} 3307

\bibitem{him3} Hall R L 1978 
                     {\it J. Phys. A} {\bf 11} 1227

\bibitem{Cohen_1985} Cohen L and Lee C 1985
                                {\it J. Math. Phys.} {\bf 26} 3105

\bibitem{IJQC_1991} Osadchii M S and Muraktanov V V 1991
                              {\it Int. J. Quant. Chem.} {\bf 39} 173

\bibitem{him4} Za\l{}uska-Kotur M A, Gajda M, Or\l{}owski A and Mostowski J 2000
                      {\it Phys. Rev. A} {\bf 61} 033613

\bibitem{him5} Yan J 2003
                      {\it J. Stat. Phys.} {\bf 113} 623

\bibitem{him6} Gajda M 2006
                      {\it Phys. Rev. A} {\bf 73} 023603

\bibitem{him7} Zhao-Liang W, An-Min W, Yang Y and Xue-Chao L 2012
                      {\it Commun. Theor. Phys.} {\bf 58} 639

\bibitem{him8} Armstrong J R, Zinner N T, Fedorov D V and Jensen A S 2011
                      {\it J. Phys. B} {\bf 44} 055303

\bibitem{him9} Ko\'scik P and Okopi\'nska A 2013
                       {\it Few-Body Syst.} {\bf 54} 1637

\bibitem{him10} Schilling C 2013
                       {\it Phys. Rev. A} {\bf 88} 042105

\bibitem{EPJD_2014} Bouvrie P A, Majtey A P, Tichy M C, Dehesa J S and Plastino A R 2014
                               {\it Eur. Phys. J. D} {\bf 68} 346

\bibitem{him11} Benavides-Riveros C L, Toranzo I V and Dehesa J S 2014
                       {\it J. Phys. B} {\bf 47} 195503

\bibitem{him12} Armstrong J R, Volosniev A G, Fedorov D V, Jensen A S and Zinner N T 2015
                            {\it J. Phys. A} {\bf 48} 085301

\bibitem{him13} Schilling C and Schilling R 2016
                       {\it Phys. Rev. A} {\bf 93} 021601(R)

\bibitem{him14} Klaiman S, Streltsov A I and Alon O E 2017
                           {\it Chem. Phys.} {\bf 482} 362

\bibitem{JPA_2017} Alon O E 2017
                            {\it J. Phys. A} {\bf 50} 295002

\bibitem{JPCS_2023} Alon O E and Cederbaum L S 2023
                              {\it J. Phys.: Conf. Ser.} {\bf 2494} 012014

\bibitem{him15} Alon O E and Cederbaum L S 2023
                       arXiv:2311.08138v1 [cond-mat.quant-gas]

\bibitem{benc1} Lode A U J, Sakmann K, Alon O E, Cederbaum L S and Streltsov A I 2023
                        {\it Phys. Rev. A} {\bf 86} 063606

\bibitem{benc2} Fasshauer E and Lode A U J 2016
                        {\it Phys. Rev. A} {\bf 93} 033635

\bibitem{benc3} L\'ev\^{e}que C and Madsen L B 2018
                        {\it J. Phys. B} {\bf 51} 155302

\bibitem{benc4} Lode A U J, L\'ev\^{e}que C, Madsen L B, Streltsov A I and Alon O E 2020
                        {\it Rev. Mod. Phys.} {\bf 92} 011001

\bibitem{benc5} Bhowmik A and Alon O E 2023
                        arXiv:2309.05240v1 [cond-mat.quant-gas]

\bibitem{arXiv_2023} Alon O E and Cederbaum L S 2023
                               to be submitted

\bibitem{RDM1} L\"owdin P O 1955
                        {\it Phys. Rev.} {\bf 97} 1474

\bibitem{RDM2} Coleman A J and Yukalov V I 2000
                        {\it Reduced Density Matrices: Coulson's Challenge 
                        (Lectures Notes in Chemistry vol 72)}
                        (Berlin: Springer)

\bibitem{GL1} Glauber R J 1963
                     {\it Phys. Rev.} {\bf 130} 2529

\bibitem{GL2} Naraschewski M and Glauber R J 1999
                     {\it Phys. Rev. A} {\bf 59} 4595

\bibitem{RDM3} Sakmann K, Streltsov A I, Alon O E and Cederbaum L S 2008
                        {\it Phys. Rev. A} {\bf 78} 023615

\end{thebibliography}
\end{document}